\documentstyle[epsfig]{prhep97}

\makeatletter
\let\chapter\hid@chapter
\makeatother
\begin{document}
\thispagestyle{empty}

\authorrunning{W.\,Ochs}
\titlerunning{{\talknumber}: Hadron and jet multiplicity }
 

\def\talknumber{403} 

\title{{\talknumber}: QCD connection between hadron and jet multiplicities}
\author{Wolfgang\, Ochs 
(wwo@mppmu.mpg.de)}
\institute{Max-Planck-Institut f\"ur Physik, F\"ohringer Ring 6,
              D-80805 Munich, Germany}
\null               

{\large             

\rightline{MPI-PhT/97-82}
\rightline{November 30th, 1997}
\rightline{talk Nr. 403}

\vfill

 \centerline{\LARGE\bf QCD connection between hadron and jet 
multiplicities\footnote{\normalsize
 to be published in the Proceedings of the
Int. Europhysics Conference on High Energy Physics,
Jerusalem, Israel, August 1997} }          

\vspace{0.7cm}      

\centerline{WOLFGANG OCHS}
\vspace{0.7cm}

\centerline{\it Max-Planck-Institut f\"ur Physik}
\centerline{\it (Werner-Heisenberg-Institut)}
\centerline{\it F\"ohringer Ring 6, D-80805 M\"unchen, Germany}


\vfill

\centerline{\bf Abstract}
\bigskip
{
\baselineskip=24pt
\noindent
The perturbative QCD provides a good overall description of both
the jet and hadron multiplicities in the $e^+e^-$ annihilation
reaction. In this
description the hadrons are considered as dual  to partons at a small
resolution parameter $Q_0$ characteristic of a hadronic scale
 of few hundred MeV.
} 
\vfill
}

\newpage

\maketitle

\begin{abstract}
The perturbative QCD provides a good overall description of both
the jet and hadron multiplicities in the $e^+e^-$ annihilation
reaction. In this
description the hadrons are considered as dual  to partons at a small
resolution parameter $Q_0$ characteristic of a hadronic scale 
 of few hundred MeV.
\end{abstract}
\section{Dual picture relating partons and hadrons}
The production rates for jets and their distribution over the
kinematic variables are described very well 
by the perturbative QCD and this success seems to be continued
by recent measurements at TEVATRON and HERA. Such a success is at
first sight astonishing as theory and experiment relate quite
different objects: a 100 GeV jet seen in the experiment consists
of a bundle of several dozens of hadronic particles whereas in the
theoretical treatment this object is often represented by only
one or two partons. Although an understanding at a fundamental
level of the colour confinement process is not yet available 
there are models which
explain how the initial partons first evolve perturbatively into a parton
jet and then by nonperturbative processes into the observed hadrons.
These nonperturbative processes look hopelessly complicated in view
of the large variety of particle and resonance species.

Despite this complication it appears that after 
averaging over some degrees
of freedom the perturbative description can 
actually be extended from jets to
single hadrons. An example is the 
good description of the hadron energy
spectrum in a jet which led to the notion of 
\lq\lq local parton hadron duality''
(LPHD) \cite{LPHD}. It is important to explore further the connection
between hadronic and partonic final states to learn about the confinement
process. Here we discuss a recent comparison \cite{lo} 
at the same but variable resolution
scale, including the 
limit where jets are fully resolved into hadrons.
\section{From jets to hadrons}
The mean jet multiplicity in 
$e^+e^-$ events can be described \cite{lo} by perturbative calculations as
function of the (\lq\lq Durham'') resolution parameter $Q_c$ 
in the full measured range down to $Q_c\sim 1$ GeV and
smoothly connects to the multiplicity of hadrons in the limit of
small $Q_c\to Q_0$; here $Q_0$ is a nonperturbative parameter
of size 250 to 500 MeV depending on the approximation scheme.
This is shown in Fig.~1
for the lower data set (\lq\lq Jets'') from LEP-1 ($Q$=91~GeV kept fixed);
the upper data set (``Hadrons'') refers to the hadronic multiplicity 
(taken as ${\cal N}=\frac{3}{2}{\cal N}_{ch}$) as function
of total energy $Q$. The only free parameters are the QCD scale 
$\Lambda$ and the hadronization scale $Q_0$ (here $\Lambda$=500 MeV 
and $\ln\frac{Q_0}{\Lambda}$=0.015).
\begin{figure}[htb]
\hspace{1.7cm}
\mbox{\epsfig{file=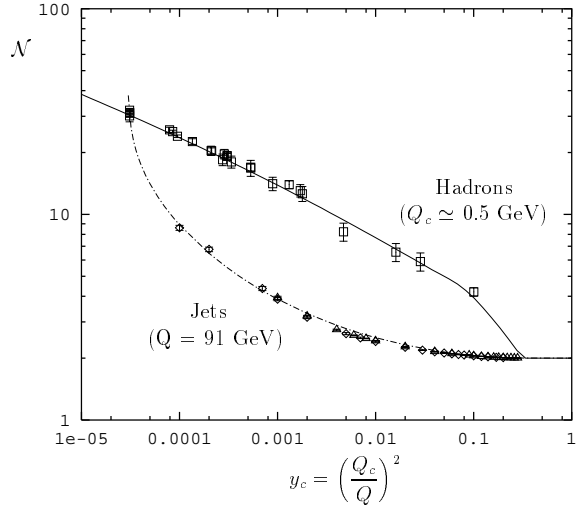,width=8.9cm,bbllx=4.cm,bblly=12.cm,bburx=21.cm,bbury=22.cm}}
\vspace{0.6cm}
\caption{Multiplicity of jets for varying resolution parameter $Q_c$
and of hadrons for varying $cms$ energy $Q$ at fixed resolution parameter
$Q_c=Q_0$ obtained from a fit to data in comparison with the perturbative
calculation.}
\vspace{-0.3cm}
\end{figure}
The curves represent the perturbative QCD calculation: two coupled evolution
equations for the multiplicities in quark and gluon jets which include
the angular ordering condition for soft gluons and energy conservation are
solved numerically  with the threshold condition ${\cal N}=2$ and then
matched with the exact $O(\alpha_s)$ result.
This improved accuracy of the solution
yields a better description than previously in regions where the coupling
becomes large, namely for hadrons near threshold and jets at high
resolution. Remarkably, both the jet and hadron multiplicities can be
described now with the same absolute normalization. 

This analysis suggests that sufficiently inclusive properties
of a jet of hadrons 
can be computed perturbatively from the jet of partons at 
all resolution scales down to the hadronization scale $Q_0$ of few
hundred MeV
and that the confinement process should be correspondingly soft.

%

\end{document}